\documentclass[aps,prl,twocolumn,groupedaddress]{revtex4}

\usepackage{graphicx}

\begin{document}
\preprint{YOSHITA/single-wire PL/Jan. 11}

\title{
Evolution of excitons via biexcitons to an electron--hole plasma without level crossing between band edge and exciton in a quantum wire
}
\author{Masahiro Yoshita}%
 \altaffiliation[Also at ]{Bell Laboratories, Lucent Technologies.}%
\affiliation{
Institute for Solid State Physics, University of Tokyo, and CREST, JST\\
5-1-5 Kashiwanoha, Kashiwa, Chiba 277-8581, Japan
}%
\author{Yuhei Hayamizu}
\affiliation{
Institute for Solid State Physics, University of Tokyo, and CREST, JST\\
5-1-5 Kashiwanoha, Kashiwa, Chiba 277-8581, Japan
}%
\author{Hidefumi Akiyama}%
 \altaffiliation[Also at ]{Bell Laboratories, Lucent Technologies.}%
\affiliation{
Institute for Solid State Physics, University of Tokyo, and CREST, JST\\
5-1-5 Kashiwanoha, Kashiwa, Chiba 277-8581, Japan
}%

\author{Loren N. Pfeiffer}
\affiliation{
Bell Laboratories, Lucent Technologies,\\
600 Mountain Avenue, Murray Hill, NJ 07974
}%
\author{Ken W. West}
\affiliation{
Bell Laboratories, Lucent Technologies,\\
600 Mountain Avenue, Murray Hill, NJ 07974
}%

\author{Kenichi Asano}
\affiliation{
Department of Physics, Osaka University, and CREST, JST\\
Toyonaka, Osaka 560-0043, Japan
}%

\author{Tetsuo Ogawa}
\affiliation{
Department of Physics, Osaka University, and CREST, JST\\
Toyonaka, Osaka 560-0043, Japan
}%

\date{1/11/2005} 

\begin{abstract}
A recent single quantum wire is of sufficient quality to reveal new details of the photoluminescence (PL) evolution with increasing electron--hole (e--h) pair density. At a pair density of 3.6 $\times$ 10$^{3}$ cm$^{-1}$, the PL is characteristic of biexcitons shifted below the exciton peak by the 2.8-meV biexciton binding. At the pair density of 1.2 $\times$ 10$^{5}$ cm$^{-1}$, the biexciton peak broadens without energy shift to an e--h plasma. At all pair densities up to 30 K, neither the exciton peak nor the one-dimensional (1D) continuum edge shows any shift. In contrast to prevailing theories, the low-energy edge of the plasma PL line never crosses the exciton peak and never makes contact with the 1D e--h continuum.

\end{abstract}


\maketitle


Many-body effects in one-dimensional (1D) electron--hole (e--h) systems where Coulomb interactions become more important have recently attracted much attention, both theoretical \cite{RossiPRL,HwangPRB,TassonePRL,PiermarocchiSSC,DasSarmaPRL} and experimental \cite{CingolaniPRB,WegscheiderPRL,AmbigapathyPRL,VouillozSSC,RubioSSC,CrottiniSSC,GuilletPRB,AkiyamaPRB}. One of the basic questions is whether the crossover from excitons to dense e--h systems is well described by the familiar picture of the exciton Mott transition.
In this picture, the binding energy of excitons is decreased with carrier density by screening, while the band edge is red-shifted because of band-gap renormalization (BGR), and the Mott transition to the e--h plasma occurs at the critical density {\it n}$_{c}$ (the Mott density) where the energy position of the band edge coincides with that of the exciton  \cite{HaugPQE}. In 2D or 3D systems, this picture is accepted as plausible description of the transition from excitons to an e--h plasma.

In 1D e--h systems, previous photoluminescence (PL) studies report the interesting experimental fact that the PL peak positions of the quantum wires are almost independent of the e--h pair densities \cite{WegscheiderPRL,AmbigapathyPRL,VouillozSSC,RubioSSC,CrottiniSSC,GuilletPRB,AkiyamaPRB}. 
This constancy of the peak position has been regarded as an exact cancellation between the shrinkage of the band gap and the reduction of the exciton binding energy on the basis of this picture \cite{DasSarmaPRL}.
On the other hand, recent intensive PL studies on 0D quantum dots \cite{DekelPRL} have shown that PL peak positions of excitons, biexcitons, and other neutral exciton complexes in 0D dots do not shift with carrier densities as a result of 3D confinement, and the same phenomena should happen in 0D localized states in wires with large localization energy or inhomogeneous broadening \cite{WegscheiderPRL,AmbigapathyPRL,VouillozSSC,RubioSSC,CrottiniSSC,GuilletPRB}. Thus, suppression of the localization effect is essential for investigations of intrinsic 1D-exciton-Mott-transition problem, which requires unprecedented high-quality quantum wires with minimized inhomogeneous broadening, at least smaller than many-body interaction energies such as a biexciton binding energy \cite{AkiyamaPRB}. 

In this Letter, we use a high-quality single T-shaped quantum wire (T wire) fabricated by the cleaved-edge overgrowth method \cite{PfeifferAPL} with a growth-interrupt annealing technique \cite{YoshitaJJAP_APL}, and study the evolution of PL shape, energy, and intensity during the crossover from dilute 1D free excitons to a dense e--h plasma. The spectral changes observed indicate that the crossover occurs {\it continuously} via {\it biexcitons}.
The position of the 1D-continuum band edge above the exciton PL is observed not to shift with the excitation power, while the plasma PL peak seen when the carrier density is high strongly broadens, so that the low-energy edge of the plasma PL shows a red shift with increasing density. However, the energy of the band edge of the plasma never crosses that of the exciton.


The single T wire consisted of a 14-nm-thick Al$_{0.07}$Ga$_{0.93}$As quantum well (QW) (stem well) grown on a (001) substrate, and an intersecting 6-nm-thick GaAs QW (arm well) overgrown on a cleaved (110) edge of the stem well. We improved the interface flatness of the arm well by performing growth-interrupt {\it in-situ} annealing at 600$^{\circ}$C for 10 minutes on the arm-well upper surface \cite{YoshitaJJAP_APL}. The sample structure and fabrication procedures are reported in detail in separate papers \cite{HayamizuAPL,YoshitaPHYE}.


\begin{figure}
\includegraphics[width=.450\textwidth]{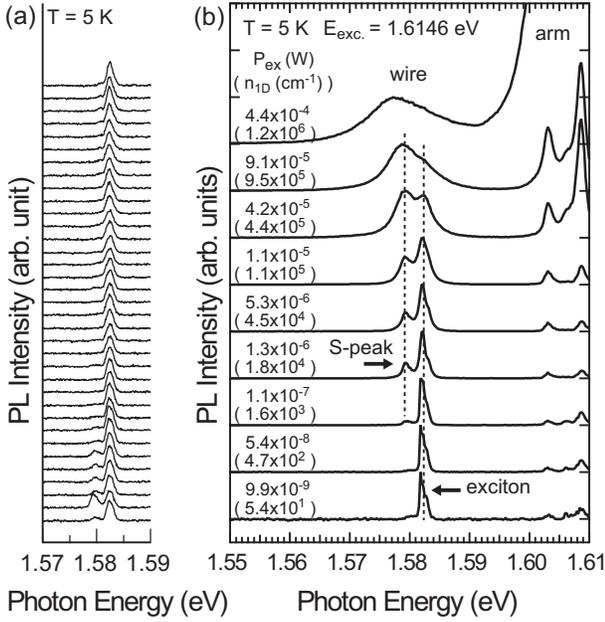}
\caption{(a) Spatially resolved PL spectra of the T wire at 5 K under point excitation scanned along the wire by steps of 0.5 $\mu$m. 
(b) Normalized PL spectra of the T wire for various excitation powers (P$_{ex}$) at 5 K. Numbers in parentheses are estimated 1D e--h pair densities. Low-energy side peak appeared below the wire exciton peak is denoted as S-peak. Two dashed vertical parallel lines are drawn to guide the eyes.}
\end{figure}

Figure 1(a) shows spatially resolved PL spectra of the wire at 5 K obtained by scanning the 0.8-$\mu$m-diameter excitation spot along the wire by steps of 0.5 $\mu$m over a length of 15 $\mu$m. The dominant PL peak at 1.582 eV has a narrow width of 1.3meV and uniform peak intensity and energy. PL excitation (PLE), absorption, and PL imaging measurements have ascribed this peak to 1D free excitons in the wire with small Stokes shift less than 0.3 meV \cite{YoshitaPHYE,AkiyamaAPL,Hayamizu2005,Takahashi2005}. 
Tiny PL peaks observed at 1.580 eV are from localized states due to monolayer islands formed on the wire during {\it in-situ} annealing \cite{YoshitaJJAP_APL} of the arm-well top surface. The spectra demonstrate that the localized states are rare while the free exciton state is continuous at least over 10 $\mu$m in the wire. 
The 1.3-meV width of the wire PL at 1.582 eV, and the wire PL shape shown in the bottom curve of Fig. 1(b), are limited by thickness inhomogeneity of the stem well, where one monolayer thickness difference gives a wire PL shift of 0.28 meV.

Excitation power dependence of PL spectra from the T wire under point excitation observed at a uniform wire region at 5 K is shown in Fig. 1(b). The energy of the excitation light was 1.6146 eV. This excitation created e--h pairs in the T wire, and the arm well, but not in the stem well. 
To eliminate influence of carrier diffusion along the wire on the PL spectra, we used a spectrograph technique. We projected the PL images of the wire by magnification of 32 onto the 25$\mu$m-wide entrance slit of the imaging spectrometer with aligning the wire parallel to the slit, and obtained PL spectra at a particular row of the 27$\mu$m-high charge-coupled-device pixels. In this way, PL spectra from 0.8$\mu$m$\times$0.8$\mu$m area on the sample centered at the excitation position was monitored. 
The e--h pair density $n_{1D}$ for each excitation power was estimated via the analysis explained later.

At excitation powers less than 10$^{-7}$ W, we observed a single PL peak at 1.582 eV due to 1D free excitons. 
As the excitation power was increased from 10$^{-7}$ to around 10$^{-5}$ W, a new PL peak (denoted hereafter as S-peak) appeared 2.8 meV below the free-exciton PL peak and increased its intensity superlinearly. Above 10$^{-5}$ W, the S-peak increased its intensity and width. Above 5 ${\times}$ 10$^{-5}$ W, the exciton peak faded into the high-energy tail of the S-peak, which finally showed a red shift and an asymmetric shape.


\begin{figure}
\includegraphics[width=.450\textwidth]{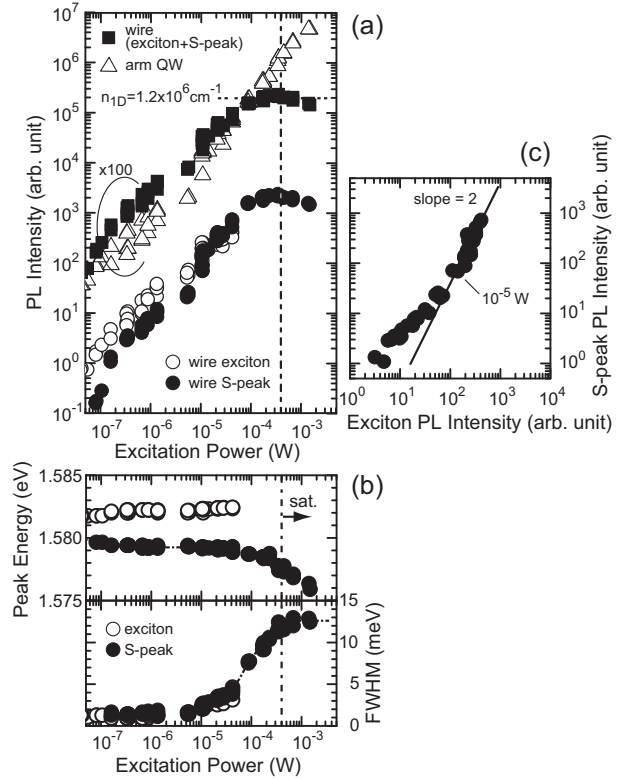}
\caption{(a) Integrated intensities and (b) positions and widths of the exciton- and S-peak in the wire PL spectra shown in Fig. 1 as a function of the excitation power. Also shown are the total PL intensity (the sum of the exciton- and S-peak intensities) of the wire and the PL intensity of the arm well. (c) Integrated PL intensity of the S-peak versus that of the exciton peak.}
\end{figure}

To investigate the PL evolution more quantitatively, we plotted, in Fig. 2, integrated area intensities (a), peak positions (b; upper panel), and widths (b; lower panel) of the exciton peak and the S-peak in the wire PL spectra. 
The total PL intensity (the sum of the exciton- and S-peak intensities) of the wire and the PL intensity of the arm well are also shown in Fig. 2(a). 
The total PL intensity of the wire becomes saturated at an excitation power of about 4 ${\times}$ 10$^{-4}$ W, while the PL intensity of the arm well still increases. This indicates that the electronic states in the wire are filled and the Fermi filling of the arm well has started. The saturation density of e--h pairs in the wire, estimated from the 21-meV energy separation between the ground states of the wire and the arm well, is 1.2 ${\times}$ 10$^{6}$ cm$^{-1}$. By assuming that the PL intensity is proportional to the e--h pair density $n_{1D}$ in the wire, we estimated $n_{1D}$ for each PL spectrum.

At excitation powers below 10$^{-7}$ W, only the exciton peak has measurable intensity. 
Above 10$^{-7}$ W, however, the S-peak appears and increases its intensity steeply (Fig. 2(a)). Figure 2(c) shows log-log plots of the integrated intensity of the S-peak against that of the exciton peak. The S-peak intensity increases superlinearly against the exciton peak intensity, and the slope gradually approaches to 2. 
The energy position and the width of the S-peak and the exciton peak are constant in the excitation power range 2 ${\times}$ 10$^{-7}$ to about 10$^{-5}$ W (Fig. 2(b)).  
Therefore, we ascribe the S-peak in this excitation range to biexcitons. 
This excitation power range corresponds to the pair density range 3.6 ${\times}$ 10$^{3}$ to 1.2 ${\times}$ 10$^{5}$ cm$^{-1}$. The corresponding mean distances $r_s$ between carriers are 220$a^*_B$ and 6.6$a^*_B$, respectively, where $a^*_B$ (=12.7 nm) is the Bohr radius of bulk GaAs. The estimated densities were reasonable values for the formation of biexcitons. 
In our previous PL studies for 20-period T wires \cite{AkiyamaPRB}, we observed similar spectral evolution. However, we could not clearly resolve biexciton formation and many other quantitative features shown below, because PL from localized states was superimposed on the PL spectra. 
The biexciton binding energy in the T wire estimated from the PL spectra was 2.8 meV in agreement with our earlier study on multiple T wires \cite{AkiyamaPRB}, but larger than those in other wires with stronger localization and/or larger size \cite{CrottiniSSC,GuilletPRB,LangbeinPRB}.

Above 10$^{-5}$ W, the width of the S-peak increases significantly from 2.0 to 12 meV with the excitation power (Fig. 2(b)). The width of the exciton peak also increases, but its intensity saturates at 5 ${\times}$ 10$^{-5}$ W and then decreases (Fig. 2(a)). In this power region, the two PL peaks are well fitted with Lorentzians, which means that homogeneous broadening dominates the PL widths. Then, above  5 ${\times}$ 10$^{-5}$ W up to  4 ${\times}$ 10$^{-4}$ W, the S-peak gradually becomes asymmetric in shape and dominates the wire PL. 
These spectral changes indicate that the pair density is increased so high that overlaps between excitons become strong and isolated excitons (and also biexcitons) are rare. That is, an e--h plasma are formed simultameously with excitons in the wire at around  10$^{-5}$ W, and the e--h system is completely changed to the 1D e--h plasma state at  5 ${\times}$ 10$^{-5}$ W. 
Therefore, we ascribe the S-peak in this excitation range to an e--h plasma. 
The estimated carrier density $n_{1D}$ in the wire is in the range from 1.2 ${\times}$ 10$^{5}$ ($r_s$ = 6.6$a^*_B$) at 10$^{-5}$ W to 1.2 ${\times}$ 10$^{6}$ cm$^{-1}$ ($r_s$ = 0.7$a^*_B$) at  4 ${\times}$ 10$^{-4}$ W and is consistent with this interpretation. 
Note that the biexciton PL changed continuously to the e--h plasma PL, and its peak position showed only small red shift less than 2 meV.


We recently studied dense 1D e--h systems correlated through the long-range Coulomb interactions theoretically, using the Tomonaga-Luttinger model \cite{AsanoJL}. In this theoretical study, we found that the 1D e--h plasma with Coulomb interactions should have strong biexcitonic correlations, that is, the plasma is strongly Coulomb-correlated plasma and may act like a "biexciton liquid". This theoretical result is consistent with the experimental observation that the biexciton PL changes to the plasma PL continuously keeping the centroid energy constant.


At the excitation powers above 4 ${\times}$ 10$^{-4}$ W, both the intensity and width of the wire PL were saturated while the peak position was still red-shifted up to 4 meV from the peak position of biexcitons (Fig. 2(b)). Reasons for this peak shift after saturation of the wire states are not clear at present, but possibly due to carrier interactions between the wire and the arm well and/or local heating of lattice temperature by extremely high photo-excitation.


\begin{figure}
\includegraphics[width=.35\textwidth]{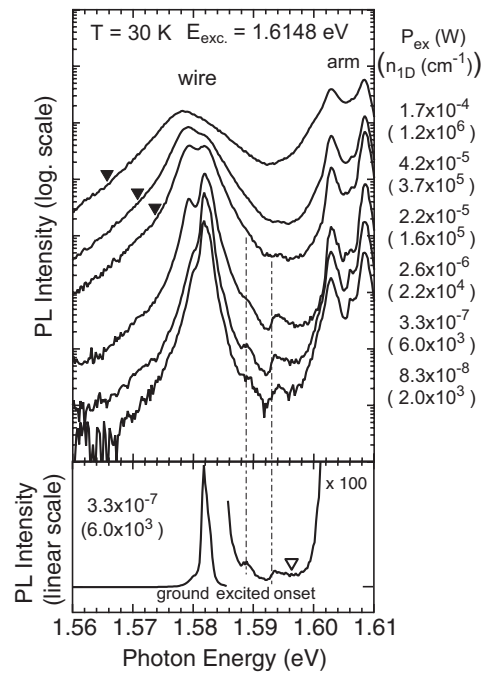}
\caption{Excitation power dependence of PL spectra from the T wire at 30 K in log. scale. The PL spectrum at 3.3 $\times$ 10$^{-7}$ W is also represented in linear scale at the bottom. Closed triangles indicate low-energy edges of the plasma PL defined as the positions where the PL intensity decreases to one twentieth of the peak intensity. The open triangle in the bottom line indicates the 1D continuum band edge \cite{AkiyamaAPL}.}
\end{figure}

Recall that all the wire PL peaks show very little shift except for the highest excitation powers. A similar feature was previously pointed out and interpreted as an exact cancellation between the shrinkage of the band gap and the reduction of exciton binding energy \cite{WegscheiderPRL,AmbigapathyPRL,GuilletPRB,DasSarmaPRL}. To verify the validity of this interpretation, we investigated PL from the band edge of the exciton continuum states observed at an elevated temperature. 

The excitation power dependence of the PL spectra from the wire at 30 K is shown in Fig. 3. At low excitation powers, not only strong PL from the exciton ground states at 1.582 eV (denoted as {\it ground}) but also a small PL peak at 1.589 eV (denoted as {\it excited}) and a continuous PL band with an onset at 1.593 eV (denoted as {\it onset}) were observed. The new small peak and the continuous band are due to an excited state of the wire excitons, and higher excited states of the excitons and 1D continuum states, respectively. These assignments were previously achieved by detailed PLE measurements and numerical calculations \cite{AkiyamaAPL}.

In the PL spectra at increased excitation powers, it is remarkable that the {\it excited} peak and the {\it onset} of the exciton excited and continuum states show no shift from their initial positions. This is still the case when the pair density is as high as 1.6 $\times$ 10$^{5}$ cm$^{-1}$ ($r_s$ = 4.9$a^*_B$) and the plasma state has already formed. At the highest pair density, 1.2 $\times$ 10$^{6}$ cm$^{-1}$ ($r_s$ = 0.7$a^*_B$), these excited-state features become smeared and buried in a tail of the broad plasma PL. However, they stay at the same energies as long as they are visible. This suggests that dense carriers contribute to scattering that broadens energy levels rather than screening that reduces exciton binding energy in the quantum wire. Thus, the absence of peak shift in Figs. 1 and 3 is not explained by an interpretation of an exact cancellation between the shrinkage of the band gap and the reduction of the exciton binding energy.

Furthermore, in Fig. 3, the low-energy edges of the plasma PL (marked by closed triangles) may reasonably be regarded as a band edge for the plasma in the wire, and these edges show large red shifts with increasing excitation powers possibly due to the BGR effect. However, they do not continuously connect to the continuum band edge (marked by an open triangle) for excitons at the lowest excitation power. Instead they begin at the energy position of biexcitons. Hence, the level crossing between the continuum band edge and the exciton expected in the exciton Mott transition never occurs. In addition, at the pair density of 1.6 $\times$ 10$^{5}$ cm$^{-1}$, the band edges for the excitons and the plasma seem to coexist. These results suggest that a new picture is necessary for the observed crossover from excitons to an e--h plasma via biexcitons in a quantum wire.

In conclusion, the evolution from excitons to an e--h plasma in a quantum wire is found to be a gradual crossover via biexcitons. The crossover from a biexciton gas to an e--h plasma occurs continuously as level broadening without peak shift. The low-energy edge of dense e--h plasma PL shows a red shift, which is regarded as BGR, but does not connect to the continuum band edge of excitons. Hence, level crossing between the band edges and the exciton never occurs. The observed crossover without such level crossing indicates no evidence for an abrupt metal--insulator phase transition, or exciton Mott transition, in the 1D e--h system.


This work was partly supported by the MEXT, Japan.


\end{document}